\documentclass[aps,prl,twocolumn,showpacs,superscriptaddress,groupedaddress]{revtex4-2}  
\usepackage[T1]{fontenc}
\usepackage{lmodern} 
\usepackage{graphicx}  
\usepackage{dcolumn}   
\usepackage{bm}        
\usepackage{amssymb}   
\usepackage{amsmath}
\usepackage{bbold}
\usepackage{array}
\usepackage{makecell}
\usepackage{braket}
\usepackage{physics}
\usepackage{titlesec} 
\usepackage[colorlinks,citecolor=blue,urlcolor=blue,hypertexnames=true]{hyperref}
\usepackage{subfigure}

\usepackage[usenames,dvipsnames,svgnames,table]{xcolor} 
\allowdisplaybreaks

\newcommand\superequiv{\mathrel{\rlap{\raisebox{\fontdimen22\textfont2}{$=$}}\raisebox{-0.5\fontdimen22\textfont2}{$ = $}}}

\begin{document}

\widetext


\title{\textcolor{Sepia}{\textbf{\Huge  Cosmological Krylov Complexity}}}


\author{ Kiran Adhikari ${}^{1}$,~{Sayantan Choudhury ${}^{2}$}}
\email{Corresponding author e-mail:
s ayantan_ccsp@sgtuniversity.org, sayanphysicsisi@gmail.com}

\affiliation{${}^{1}$RWTH Aachen University, D-52056, Aachen, Germany}
\affiliation{ ${}^{2}$Centre For Cosmology and Science Popularization (CCSP),
SGT University, Gurugram, Delhi- NCR, Haryana- 122505, India,}
\affiliation{${}^{3}$International Centre for Theoretical Sciences, Tata Institute of Fundamental Research (ICTS-TIFR), Shivakote, Bengaluru 560089, India,}

\begin{abstract}
In this paper, we study the Krylov complexity ($K$) from the planar/inflationary patch of the de Sitter space using the two mode squeezed state formalism in the presence of an effective field having sound speed $c_s$. From our analysis, we obtain the explicit behavior of Krylov complexity ($K$) and lancoz coefficients ($b_n$) with respect to the conformal time scale and scale factor in the presence of effective sound speed $c_s$. Since lancoz coefficients ($b_n$) grow linearly with integer $n$, this suggests that universe acts like a chaotic system during this period. We also obtain the corresponding Lyapunov exponent $\lambda$ in presence of effective sound speed $c_s$. We show that the Krylov complexity ($K$) for this system is equal to average particle numbers suggesting it's relation to the volume. Finally, we give a comparison of Krylov complexity ($K$) with entanglement entropy (Von-Neumann) where we found that there is a large difference between Krylov complexity ($K$) and entanglement entropy for large values of squeezing amplitude. This suggests that Krylov complexity ($K$) can be a significant probe for studying the dynamics of the cosmological system even after the saturation of entanglement entropy. 
\end{abstract}

\maketitle

\section{\textcolor{Sepia}{\textbf{ \Large Introduction}}} \label{sec:introduction}
Recent years have seen a fair amount of applications coming from quantum information theory into high energy physics and cosmology. \cite{Chapman:2021jbh,Chapman:2021eyy,Faulkner:2022mlp,Shaghoulian:2022fop,Bhattacharyya:2021ypq,Adhikari:2021ckk}. One such concept is complexity and chaos \cite{Brown:2017jil,Susskind:2014moa,Brown:2015lvg,Maldacena:2015waa}. Complexity characterizes the notion of difficulty of preparing a state or applying a certain unitary operator while chaos quantifies the sensitivity of the system to the initial condition. While in classical mechanics chaos is a very well-defined quantity, it is not so in quantum mechanics. So, one resorts to various kinds of probes and measures. One recent tool that has been proposed to study operator growth and characterize quantum chaos is Krylov/K complexity \cite{Caputa:2021sib,Parker:2018yvk,Roberts:2018mnp,Rabinovici:2020ryf,Barbon:2019wsy,Jian:2020qpp,Dymarsky:2019elm,Dymarsky:2021bjq, Adhikari:2022whf}. 

In this framework, one tries to understand the Heisenberg evolution of some initial Hermitian operators. Depending on the Hamiltonian and initial operator, the evolution can become extremely complicated. Krylov complexity can capture this growth of the operator. While obtaining Krylov complexity, one also has to obtain the so-called Krylov basis using the Lancoz algorithm. Lancoz algorithm also gives us Lancoz coefficients which are conjectured to be maximum for chaotic systems \cite{Parker:2018yvk}. In recent years, Krylov complexity has been studied extensively from black hole physics to conformal field theories. 

The main motivation to study complexity in high energy physics comes from holographic conjectures of complexity. In particular, Susskind et al \cite{Susskind:2014moa, Brown:2015lvg} conjectured that complexity can be used to probe the physics behind the black hole horizons via “complexity = volume” and “complexity = action” proposals. Following these works, circuit complexity has been computed using different techniques and even computed in the context of quantum field theory and cosmology \cite{Chapman:2021jbh,Jefferson:2017sdb,Bhattacharyya:2020rpy,Bhattacharyya:2020kgu}. However, one issue with these calculations is that the notion of complexity is very ambiguous depending on the choices of gates, reference and target gates, and arbitrary tolerance. Fortunately, Krylov's complexity is free of such choices, therefore, making it an ideal candidate to study in holographic and QFT settings. 

Our goal in this paper is to study Krylov complexity and chaos in de Sitter Cosmology with effective sound speed, and gain quantum information theoretic insights about cosmological evolution and structure formation. The reason to include the sound speed is to make our calculations as general as possible. While holography is mostly studied on the Anti de Sitter background, we seem to live in a de Sitter one. This also gives us a strong motivation to see if those holographic conjectures holds for de Sitter case too \cite{Reynolds:2017lwq,Chapman:2021eyy,An:2019opz}. Particularly, we study the Krylov complexity and chaos on the scalar cosmological perturbations on an expanding Friedmann-Lemaitre-Robertson-Walker (FLRW) background.
Scalar perturbations on an expanding background can naturally be described by the two mode squeezing operator. Modes inside the horizon are frozen while mode exiting the horizon are highly squeezed. After obtaining expressions for K-complexity and chaos for general perturbations, we apply to a simple model of de Sitter expansion where we obtain explicit expressions. 

Squeezed states and squeezing operator is an extremely important subject in quantum optics with applications from quantum computing, quantum cryptography and even in gravitational physics
\cite{PhysRevA.31.3068,ncsqo,ZELAYA20183369,PhysRevLett.97.110501,PhysRevA.103.062405,PhysRevA.61.022309,1981PhRvD..23.1693C,RevModPhys.77.513,contivar,braunstein2005quantum,PhysRevLett.117.110801,RevModPhys.58.1001,Adhikari:2021ked,Ando:2020kdz,Martin:2021znx,Vahlbruch_2007,PhysRevLett.104.103602,advinqm,doi:10.1080/00107510802091298,Xu:19,atomchp}.
  For review on the fundamentals of squeezed states, we refer to vast literautres on \cite{Ph:31,ssol,book,sql,30yrof,RosasOrtiz2019CoherentAS,Schumaker:1985zz,Garcia-Chung:2020gxy,PhysRevLett.97.011101,Chua_2014,Choudhury:2013jya}. From the cosmological point, the concept of squeezed states was introduced by Grishchuk and Sidorov \cite{Grishchuk:1990cm,Grishchuk:1991cm} on the inflationary cosmology where they analysed the features of relic gravitons and phenomena such as particle creation and black-hole evaporation. Andreas Albrecht et al. \cite{Albrecht:1992kf} also used the two-mode squeezed state formalism to understand the inflationary cosmology and the amplification process of quantum fluctuations during the inflation period. For applications of squeezed state formalism in High energy physics and in cosmology see   \cite{Hasebe:2019ibg,Choudhury:2011jt,Choudhury:2012yh,Choudhury:2011sq,Choudhury:2013zna,Choudhury:2015hvr,Bhattacharyya:2020kgu,Choudhury:2017cos,Akhtar:2019qdn,Choudhury:2016pfr,Choudhury:2016cso,Bhattacharyya:2020rpy,Choudhury:2017bou,Einhorn:2003xb,Choudhury:2017qyl,Baumann:2014nda,Grain:2019vnq,Grishchuk:1992tw,Bhargava:2020fhl,Choudhury:2020hil,Adhikari:2021pvv,Choudhury:2021brg,Martin:2021qkg}. 

The structure of the paper is as follows:
\begin{itemize}
    \item In Section \ref{sec:E2}, we give a review of Krylov complexity. We review how Krylov complexity and Lancoz coefficients can be computed using Lancoz algorithms. Furthermore, we will see that Lancoz coefficients can characterize the chaotic properties of the system. For systems with symmetry, the computations of Lancoz coefficients and Krylov complexity can be simplified significantly. 
    \item In Section \ref{Cosmo}, we give a review of cosmological perturbations and obtain the quadratic Hamiltonian. In order to make the calculations as general as possible, we also include the effective sound speed $c_s$. With this quadratic hamiltonian, we obtain two-mode squeezed state formalism in section \ref{squeeze}. We obtain a set of differential equations for squeezing parameters $r_k$ and $\phi_k$. In Section \ref{complexityChaos}, we obtain the expression for Krylov complexity and Chaos for two-mode squeezing operator. We obtain the Krylov complexity to be $\sinh^2{r_k}$ and lancoz coefficients $\alpha n$. Lancoz coefficients grow linearly with $n$ showing that the system is chaotic in nature. 
    \item In Section \ref{sec:Desitter}, we apply these calculations to the de Sitter background for diferent effective sound speed. 
    \item In Section \ref{sec:Conclusion}, we give the conclusion of our work and give up future prospects.
    
\end{itemize}

\section{\textcolor{Sepia}{\textbf{ \Large Review of Krylov complexity}}}
\label{sec:E2}
In this section, we will give a brief overview of operator growth and Krylov complexity. There are different notions of complexity in literature. One approach getting popular in the high energy physics section is Nielsen's geometric approach of complexity \cite{Nielsen1, Nielsen2, Nielsen3, Nielsen4}. The interest in Krylov complexity has its origins in the certain shortcomings of Nielsen's measure. Particularly, Nielsen's complexity measure is dependent on the choice of gates, choice of reference, and target states and tolerance. This makes it very difficult to define it properly in the context of QFT or holography. In contrast, Krylov complexity is well defined and is independent of these choices. These features make it well suited for application to QFTs and holography. Furthermore, Krylov complexity and Lancoz coefficients obtained from Krylov complexity can be used to characterize the chaotic systems. For a detailed overview of Krylov complexity, we would like to refer to \cite{Caputa:2021sib}.

Consider a quantum Hamiltonian $H$ and time-dependent Heisenberg operator $\mathcal{O}(t)$. The time evolved operator is described by the Heisenberg equation:
\begin{equation}
    \partial_t\mathcal{O}(t) = i[H,\mathcal{O}(t)]
\end{equation}
where, $[A,B] = AB -BA$ is the commutator. Denoting $\mathcal{O}(0) =\mathcal{O} $, the formal solution of Heisenberg equation is given by
\begin{equation}
    \mathcal{O}(t)  =  e^{iHt}\mathcal{O}e^{-iHt}
\end{equation}
Using the Baker-Campbell-Hausdorff (BCH) formula
\begin{equation}
    e^XYe^{-X} = \sum_{n=0}^\infty \frac{\mathcal{L}_X^n Y}{n!}
\end{equation}
where $\mathcal{L}_X$ is the Liouvillian super-operator defined as $\mathcal{L}_X Y = [X,Y]$, we can obtain the time evolution series for $\mathcal{O}(t) $ as:
\begin{equation}
\begin{aligned}
 \mathcal{O}(t)  &=  \sum_{n=0}^\infty \frac{(it)^n}{n!}\mathcal{L}_H^n \mathcal{O} \\
 &= \mathcal{O} + it [H,\mathcal{O} ] + \frac{(it)^2}{2!}[H,[H, \mathcal{O}]]\\
 & + \frac{(it)^3}{3!}[H,[H,[H,\mathcal{O}]]] + \dots
\end{aligned}
\end{equation}
With the time evolution the spreading of initial operators occurs and this means more complicated nested commutators need to be accounted. This gives a notion of complexity of the Heisenberg operator as a function of time. Krylov complexity quantifies this growth in a precise manner. In general, if the Hamiltonian is chaotic, the nested commutators for the operator $\mathcal{O}$ will be given by increasingly complex operators. 

In following, we will drop the subscript on Liouvillian super-operator $\mathcal{L}_H$ as we will be only focusing on hamiltonian $H$ and represent the repeated action of $\mathcal{L}$ as $\tilde{\mathcal{O}}_n =\mathcal{L}^n \mathcal{O}$. Then, the time evolution series for $\mathcal{O}(t) $ can be written as:
\begin{equation}
    \label{eq:timeSeries}
    \mathcal{O}(t) = e^{i\mathcal{L}t}\mathcal{O} = \sum_{n=0}^\infty \frac{(it)^n}{n!}\mathcal{L}^n \mathcal{O} = \sum_{n=0}^\infty \frac{(it)^n}{n!}\tilde{\mathcal{O}}_n
\end{equation}
We can now interpret (\ref{eq:timeSeries}) as the Schrodinger's time evolution where $\mathcal{O}(t)$ plays the role of "operator's wave functions", and the Liouvillian $\mathcal{L}$ as the Hamiltonian. Then we associate $|\mathcal{O}) $ with the Hilbert space vectors corresponding to the operator $\mathcal{O}$ as:
\begin{equation}
    \mathcal{O} \superequiv |\mathcal{\tilde{O}}), \mathcal{L}^1\mathcal{O} \superequiv |\mathcal{\tilde{O}}_1), \mathcal{L}^2\mathcal{O} \superequiv |\mathcal{\tilde{O}}_2),
    \mathcal{L}^3\mathcal{O} \superequiv |\mathcal{\tilde{O}}_3),\dots
\end{equation}
It is not necessary that these operators form an orthonormal basis a prior. However, starting from these basis $|\mathcal{\tilde{O}}_n)$, we can use a version of 
Gram–Schmidt orthogonalization procedure called Lanczos algorithm to construct orthonormal basis, known as Krylov basis 
$|\mathcal{O}_n)$. For this, we need a choice of inner product of these operators. One natural choice is Wightman norm:
\begin{equation}
\label{eq:wightmanNorm}
    (A|B) = \langle e^{H\beta/2}A^\dagger e^{-H\beta/2} B \rangle_\beta
\end{equation}
where $\langle \dots \rangle _\beta = \Tr{e^{-\beta H}\dots} /\Tr{e^{-\beta H}}$ is the thermal expectation value at temperature $1 / \beta$.

In order to obtain the Krylov basis using the Lancoz algorithm, we can use the fact that first two operators in $|\mathcal{\tilde{O}}_n)$ are orthogonal with respect to (\ref{eq:wightmanNorm}). So, we can include them in Krylov basis as:
\begin{equation}
   |\mathcal{O}_0) :=   |\mathcal{\tilde{O}}_0) = |\mathcal{O}), |\mathcal{O}_1) := b_1^{-1} \mathcal{L} |\mathcal{\tilde{O}}_0)
\end{equation}
where $b_1 = \sqrt{(\mathcal{\tilde{O}}_0\mathcal{L}|\mathcal{L}\mathcal{\tilde{O}}_0)}$ normalized the vector. Then, we can construct the next states iteratively as:
\begin{equation}
    |A_n) = \mathcal{L}|\mathcal{O}_{n-1})-b_{n-1}|\mathcal{O}_{n-2})
\end{equation}
followed by normalization:
\begin{equation}
    |\mathcal{O}_n) = b_n^{-1}|A_n), b_n = \sqrt{(A_n|A_n)}
\end{equation}
We need to run this algorithm until $b_n$ hits zero, then in additon to a full orthonomal basis called Krylov basis as well  coefficients $b_n$ which are called Lancoz coefficients. These lancoz coefficients are extremely useful and characterize the chaos of the system.

Once we obtain the Krylov basis, we can represent the time evolved operator $\mathcal{O}(t)$ as:
\begin{equation}
\label{eq:schrodinger}
    |\mathcal{O}(t)) = e^{i\mathcal{L}t}|\mathcal{O}) = \sum_n i^n \phi_n(t)  |\mathcal{O}_n)
\end{equation}
where the amplitudes $\phi_n(t)$ are real, and $|\phi_n|^2$ can be thought up of probablities which sums up to one. 
\begin{equation}
    \sum_{n}|\phi_n|^2 =  \sum_{n}p_n =  1
\end{equation}
In order to obtain these amplitudes, we can think of (\ref{eq:schrodinger}) as a form of Schrodinger equation. Then, we take the partial derivative on (\ref{eq:schrodinger})  to obtain

\begin{equation}
\label{eq:partial}
    \partial_t|\mathcal{O}(t)) =i \mathcal{L}|\mathcal{O}(t)) =  \sum_n i^n \partial_t\phi_n(t)  |\mathcal{O}_n)
\end{equation}
Applying the action of Liouvillian on Krylov basis 
\begin{equation}
\label{eq:Liouaction}
    \mathcal{L}|\mathcal{O}_{n} )= b_n|\mathcal{O}_{n-1} ) + b_{n+1}|\mathcal{O}_{n+1} )
\end{equation}
on (\ref{eq:partial}), we obtain
\begin{equation}
    \partial_t|\mathcal{O}(t)) = \sum_n i^n (b_n \phi _ {n-1} - b_{n+1}\phi _ {n+1})  |\mathcal{O}_n)
\end{equation}
Now, matching the coefficients appropriately on (\ref{eq:partial}), we obtain the time evolution of amplitude as

\begin{equation}
    \partial_t \phi_{n}(t)  = b_n \phi _ {n-1} - b_{n+1}\phi _ {n+1}
\end{equation}
This equation can be solved with the knowledge of Lanczos coefficients $b_n$
and with initial condition $\phi_n(0) = \delta _{n0}$.
Once we have obtained the expression for amplitudes $\phi_n$, we can then finally give the expression for Krylov complexity.

Krylov complexity/K-complexity is given by:
\begin{equation}
    K = \sum_n n |\phi_n|^2
\end{equation}
One crucial advantage of the Lancoz algorithm we discussed above is that it also has a potential to capute the chaotic properties of the system. 
It was conjecture that Lancoz coefficients in a quantum system is bounded linearly as
\begin{equation}
    b_n \leq \alpha n + \gamma 
\end{equation}
where $\alpha$ is the operator growth rate and $\gamma$ is the constant depending on the operator. These two parameters are usally obtained from the Hamiltonian of the system. For extremely chaotic system, Krylov complexity grows exponentially fast with an exponent $\gamma$ given by:
\begin{equation}
    \gamma = 2 \alpha
\end{equation}
In literature, this exponent has also been associated with Lyapunov exponent. For example, at finite temperature $T = 1/\beta$, one obtains $\alpha = \pi / \beta$. In \cite{Maldacena:2015waa}, it was conjectured that this bound the lyapunov exponent, i.e. maximal chaotic system.

In a certain class of systems which enjoys symmetry, Krylov complexity can be computed analytically using the techniques developed in \cite{Caputa:2021sib}. For these systems with symmetry group, action of Liouvillian $\mathcal{L}$ on the Krylov basis can be seen as an action of raising and lowering operators
\begin{equation}
\label{eq:ladderOperation}
   \mathcal{L} = \alpha (L_+ + L_-) 
\end{equation}
  The parameter $\alpha$ is dependent on the system we are considering, and influence the chaotic properties of the system. Furthermore, we can read off the Lancoz coefficients immediately with this approach from the action of ladder operators on Krylov basis.
 \begin{equation}
 \label{eq:actionOnBasis}
     \alpha L_+|\mathcal{O}_{n} ) = b_{n+1}|\mathcal{O}_{n+1} ),  \alpha L_-|\mathcal{O}_{n} ) = b_{n}|\mathcal{O}_{n-1} )
 \end{equation}
 There are several examples of symmetries explored in \cite{Caputa:2021sib} such as $SL(2,R), SU(2)$. For us, $SL(2,R)$ will be the most relevant.

\section{\textcolor{Sepia}{\textbf{ \Large  Krylov complexity of Cosmological perturbations}}}
\label{Cosmo}
Now that we have explored the concept of Krylov complexity and lancoz coefficients,
 we will now apply it to the scalar cosmological perturbations. The operator of interest for our case is two mode squeezing operator. For a detailed review of cosmological quantum perturbations and quantum fields in curved space time, we refer to \cite{MUKHANOV1992203,Mukhanov:2007zz}. 

 We consider a spatially flat Friedmann-Lemaitre-Robertson-Walker (FLRW) metric:
\begin{equation}
    ds^2 = -dt^2 + a(t)^2d\Vec{x}^2 = a(\tau)^2(-d\tau^2+d\Vec{x}^2)
\end{equation}
We will consider linear scalar field fluctuation $\varphi(x) = \varphi_0(t) + \delta\varphi(x)$ on this background metric to obtain the metric
\begin{equation}
    ds^2 = a(\tau)^2 ( - (1+2\psi(x,\tau))d\tau^2 + (1-2\psi(x,\tau))d\Vec{x}^2) 
\end{equation}
We will define a curvature perturbation term $R = \psi + \frac{H}{\Dot{\varphi}_0}\delta\varphi$. Here, dot indicates derivative with respect to cosmic time $t$, and $H = \frac{\Dot{a}}{a}$. If we insert these conditions into the total action and expand to second order, the action becomes \cite{MUKHANOV1992203}:
\begin{equation}
\label{eq:cosmoAction}
    S = \frac{1}{2}\int dtd^3xa^3\frac{\varphi_0^2}{ c_s^2 H^2}\left[ \Dot{R}^2- c_s^2 \frac{1}{a^2}(\partial_iR)^2 \right]
\end{equation}
where $c_s= \sqrt{\Dot{p}/\Dot{\rho}}$ the effective sound speed of the effective fluid. Here, $p$ and $\rho$ corresponds to the effective pressure and density of the effective fluid. For more details on effective sound speed, we refer to the literature on \cite{Albrecht:1992kf}. The effective sound speed is bounded by one to maintain the causality. Cosmological observations restrict the lower bound at $c_s = 0.024$ \cite{Choudhury:2021brg}. So, we obtain the bound to be $0.024 \leq c_s \leq 1$. Physically, $c_s =1$ describes a single scalar field slow roll model while $c_s <1$ describes a wide class of non-canonical scalar field theories. 

If we instead expand up to the third order, we will get non-gaussian terms \cite{Maldacena:2002vr} too, which are also very interesting to study. For our purpose, we will restrict to second order i.e. only up to gaussian states.

We will define Mukhanov variable $\nu \equiv zR$, where $z \equiv \frac{a\sqrt{2}\epsilon}{c_s} $, with $\epsilon = -\frac{\Dot{H}}{H^2} = 1- \frac{H'}{H^2}$. Here, the prime $'$ indicates derivative with respect to conformal time. With this, the action (\ref{eq:cosmoAction}) becomes:
\begin{equation}
    S = \frac{1}{2}\int d\tau d^3x\left[ \nu'^2 - \frac{c_s^2}{a^2} (\partial_i \nu)^2 + \frac{z''}{z}\nu^2 \right]
\end{equation}
Each mode will evolves independently. These modes satisfy the harmonic oscillator equation with time dependent effective mass from time dependence of the background. We can then quantize this harmonic oscillator according to the standard quantization technique of the harmonic oscillator.
So, we will promote these perturbations to quantum fields and expand them to fourier series.
\begin{equation}
    \hat{\nu}(\tau, \Vec{x}) = \int \frac{d^3k}{(2\pi)^{3}} \hat{\nu}_k(\tau)\exp(i\Vec{k}.\Vec{x})
\end{equation}
We then define the usual creation and annihilation operators:
\begin{equation}
    \hat{v}_k = \frac{1}{\sqrt{2k}}(\hat{a}_k + \hat{a}_{-k})\text{  ,  } \hat{v}_k' = -i\frac{k}{2}(\hat{a}_k + \hat{a}_{-k})
\end{equation}
 With this, the quadratic hamiltonian  becomes:
 \begin{align}
 \label{eq:quadHamiltonian}
     \hat{H} &= \frac{1}{2}\int d^3k\hat{\mathcal{H}}_k \nonumber\\
     &= \frac{1}{2}\int d^3k \left[ \Omega_k(\hat{a}_k\hat{a}_k^\dagger + \hat{a}_{-k}^\dagger\hat{a}_{-k}) - i\beta_k( \hat{a}_k\hat{a}_{-k}- \hat{a}_k^\dagger \hat{a}_{-k}^\dagger)  \right] \nonumber\\
     &
 \end{align}
 where,
 \begin{align}
     \label{eq:parameters}
     \Omega_k &= \frac{k}{2}(1+c_s^2) \nonumber \\
     \beta_k &= \sqrt{\left(\frac{k}{2}(1-c_s^2) \right)^2+ \left(\frac{z'}{z} \right)^2} 
 \end{align}
 \subsection{\textcolor{Sepia}{\textbf{ \Large Squeezed states formalism}}}
 \label{squeeze}
The first term in the hamiltonian (\ref{eq:quadHamiltonian}) represents free particle hamiltonian. The second term shows the interaction between the quantum perturbation and the expanding background. Given this quadratic hamiltonian $\hat{\mathcal{H}}_k$, the unitary evolution $\mathcal{U}_k$ can be factorized into product of two mode rotation operator $\hat{R}_k(\beta_k)$ and two mode squeezing operator $\hat{S}_k(r_k,\phi_k)$ \cite{Albrecht:1992kf}:
 \begin{equation}
     \mathcal{U}_k = \hat{S}_k(r_k,\phi_k)\hat{R}_k(\beta_k)
 \end{equation}
 The two mode rotation operator $\hat{R}_k(\beta_k)$ in terms of rotational parameter is given by:
 \begin{equation}
     \hat{R}_k(\beta_k) = \exp \left[ -i\beta_k(\tau)(\hat{a}_k\hat{a}_{k}^\dagger + \hat{a}_{-k}^\dagger \hat{a}_{-k} ) \right]
 \end{equation}
 while two-mode squeeze operator $\hat{S}_k(r_k,\phi_k)$ in terms of squeezing parameter $r_k(\tau)$ and squeezing angle $\phi_k$ is given by:
 \begin{align}
 \label{eq:cosmologicalSqueezeOperator}
     \hat{S}_k(r_k,\phi_k) &= \exp \left[ \frac{r_k(\tau)}{2}\left( e^{-2i\phi_k(\tau)}\hat{a}_k\hat{a}_{-k}-  e^{2i\phi_k(\tau)}\hat{a}_{-k}^\dagger\hat{a}_k\dagger \right)\right] \nonumber \\
     & 
 \end{align}
 Since the rotation operator only changes the phase, we will ignore the rotation operator hereon as it doesn't have much consequences. When the two-mode squeezing operator acts on the vacuum, it gives squeezed vacuum states
 \begin{align}
     \ket{SQ(k,\tau)} &= \hat{S}_k(r_k,\phi_k) \ket{0_k,0_{-k}} \nonumber\\
     &= \frac{1}{\cosh r_k}\sum_{n= 0}^\infty e^{-2in\phi_k}\tanh^n r_k\ket{n_k,n_{-k}}
 \end{align}
 where,
 \begin{equation}
     \ket{n_k,n_{-k}} = \left[ \frac{1}{n!}(a_k^\dagger a_{-k}^\dagger)^n \right]\ket{0_k,0_{-k}}
 \end{equation}
 The two-mode squeezed vacuum is normalized: 
 
 \begin{align}
    & \langle SQ(k,\tau) | SQ(k,\tau) \rangle \nonumber  \\
    &= \frac{1}{\cosh^2r_k} \sum_{n= 0}^\infty\sum_{m= 0}^\infty e^{-2i(n-m)\phi_k}\tanh^{m+n}r_k\delta_{m,n} \nonumber\\
    &= \frac{1}{\cosh^2r_k} \sum_{n= 0}^\infty \tanh^{2n}r_k = 1
 \end{align}
 The full wave function corresponding to all modes can be obtained straightforwardly as a tensor product of each $k$:
 \begin{equation}
     \ket{SQ(\tau)} = \otimes_{k} \ket{SQ(k,\tau)}
 \end{equation}
 One can obtain the time evolution of the squeezing parameters $r_k(\tau),\phi_k(\tau)$ via Schrödinger equation:
 \begin{equation}
     i\frac{d}{d\tau}\ket{SQ(k,\tau)} = \Vec{\mathcal{H}} \ket{SQ(k,\tau)}
 \end{equation}
  This gives us a set of differential equations:
 \begin{align}
 \label{eq:evolutions}
     \frac{dr_k}{d\tau} &= - \beta_k\cos(2(\tilde{\phi_k}-\phi_k)) \\
     \frac{d\phi_k}{d\tau} &= \Omega_k  +  \beta_k\coth(2r_k)\sin(2(\tilde{\phi_k}-\phi_k))
 \end{align}
 where,
 \begin{equation}
     \tilde{\phi_k} = - \frac{\pi}{2} +\frac{1}{2} \tan^{-1}\left[ \frac{\Omega_k}{2} \left(\frac{z'}{z} \right)\left( \frac{1-c_s^2}{1+c_s^2} \right) \right]
 \end{equation}
 For a stationary background where $z$ is constant, no squeezing occurs at $r=0$. These set of differential equations (\ref{eq:evolutions}) can be solved for particular scale factor $a(\tau)$ and obtain the solution for squeezing parameters  $r_k(\tau),\phi_k(\tau)$. However, it is not always the case that nice analytical expression can be obtained, and one will have to rely on numerical methods. For an exponentially expanding de Sitter background with $c_s = 1$, such analytical expression exists and this makes it easier to study concepts like Krylov complexity and chaos. We will look at it in detail on section \ref{sec:Desitter}. For other sound speed, we will rely on numerical tools. 
 
\subsection{\textcolor{Sepia}{\textbf{ \Large Complexity and chaos}}} 
\label{complexityChaos}
The hamiltonian of interest from (\ref{eq:quadHamiltonian}) is
 \begin{align}
     \hat{H}_k 
     &= \frac{- \beta_k}{2}( \hat{a}_k\hat{a}_{-k}- \hat{a}_k^\dagger \hat{a}_{-k}^\dagger)
 \end{align}
 
 Comparing with the $SL(2,R)$ Liouvillian operator (\ref{eq:ladderOperation}), $\mathcal{L} = \alpha(L_+ + L_-)$, we can associate
 \begin{equation}
     \alpha =\frac{- \beta_k}{2}, L_+ = \hat{a}_k^\dagger \hat{a}_{-k}^\dagger, L_- = \hat{a}_k\hat{a}_{-k}
 \end{equation}
 
For the two mode squeezing operator, the Krylov basis is the standard two-oscillator Fock space
 \begin{equation}
     |\mathcal{O}_{n} )  =    \ket{n_k,n_{-k}} = \left[ \frac{1}{n!}(a_k^\dagger a_{-k}^\dagger)^n \right]\ket{0_k,0_{-k}}
 \end{equation}
 
 The Lancoz coefficients can be computed by the action of ladder operators on Krylov basis as $\alpha L_-|\mathcal{O}_{n} ) = b_{n}|\mathcal{O}_{n-1})$.
Since $\hat{a}_k\ket{n_k} = \sqrt{n} \ket{(n-1)_k}$ we obtain lancoz coefficients as
\begin{equation}
    b_n = \alpha n
\end{equation}
These lancoz coefficients grow linearly with $n$ showing that this system is chaotic in nature. Exact chaotic nature depends also on the coefficient $\alpha$ coming from the Hamiltonian (\ref{eq:quadHamiltonian}). 

Then in the Krylov basis, we can write the Heisenberg's operator state as
 
  \begin{align}
     |\mathcal{O}(t))&=  \sum_n i^n \phi_n(t)  |\mathcal{O}_n) \nonumber\\
     &= \frac{1}{\cosh r_k}\sum_{n= 0}^\infty e^{-2in\phi_k}\tanh^n r_k\ket{n_k,n_{-k}}
 \end{align}
 So the operators wave function are given by
\begin{equation}
  \phi_n =   \frac{e^{-2in\phi_k}\tanh^n r_k}{\cosh r_k}
\end{equation}
and they sum to 1 as:
\begin{equation}
 \sum_{n=0}^\infty  |\phi_n|^2 =   \frac{1}{\cosh^2r_k} \sum_{n= 0}^\infty \tanh^{2n}r_k = 1
\end{equation}
Using the operator wave function, we can now compute the Krylov complexity
\begin{align}
    K &= \sum_n   |\phi_n|^2  \nonumber \\
    &= \sum_{n= 0}^\infty n \frac{\tanh^{2n}r_k}{\cosh^2r_k} = \sinh^2r_k
\end{align}
where, we used the identity
\begin{equation}
    \sum_{m=0}^\infty m z^m = z/(1-z)^2
\end{equation}
 for $|z| \leq 0 $.
Indeed, Krylov complexity for our model saturates the bound of maximum complexity growth proposed in \cite{Hornedal:2022pkc}. The reason for this is two mode squeezed states satisfies the $SL(2,R)$ symmetry group structure, and it is argued in \cite{Hornedal:2022pkc} that this group structure belonging to generalized coherent states has maximal complexity growth.

Finally for the two mode squeezed formalism of cosmological perturbations, we can give an explicit expression of Lancoz coefficients $b_n$ and Krylov complexity $K$ as:
\begin{equation}
\label{eq:complexityChaos}
    b_n = \abs{\frac{- \beta_k}{2}}n, K = \sinh^2r_k
\end{equation}
The Krylov complexity depending on the squeezing parameters $r_k$ can now be obtained by solving the differential equations (\ref{eq:evolutions}). For low amount of squeezing i.e. $r \ll 1$, we see that the Krylov complexity $K \approx 0$. This makes sense as for low amount of squeezing, the evolved operator would be similar to the initial operator thus demonstrating an operational reasoning to the concept of Krylov complexity.

For this type of Lanczos coefficients growing linearly with $n$ in (\ref{eq:complexityChaos}), the Krylov complexity grows exponentially fast in $r_k$ with an exponent, which can also be interpreted as a Lyapunov exponent, 
\begin{equation}
 \lambda = 2 \alpha = \abs{\beta_k } = \sqrt{\left(\frac{k}{2}(1-c_s^2) \right)^2+ \left(\frac{z'}{z} \right)^2} ,
\end{equation}
 
The lyapunov exponent has a interesting structure. For $c_s = 1 $, the lyapunov exponent is just $\lambda = \abs{\frac{z'}{z}}$, and is independent of mode vectors $k$. However for different effective sound speed $c_s$ than 1, the lyapunov exponent is dependent on the mode vectors $k$ too.

The linear growth of Lancoz coefficients in (\ref{eq:complexityChaos}) indicates that this system is chaotic in nature. The origin of chaos comes from the fact that scalar cosmological perturbation behaves like an inverted harmonic oscillator at large scales. Since inverted harmonic oscillator are chaotic in nature, this feature is reflected on the expression for Lancoz coefficients and Krylov complexity in (\ref{eq:complexityChaos}).

Interestingly, the expression for Krylov complexity obtained in (\ref{eq:complexityChaos}) is equal to the average particle number in each mode:
\begin{equation}
    \langle \hat{n}_k \rangle = \langle \hat{n}_{-k}\rangle =  \sinh^2r_k = K
\end{equation}
Since volume of a system, $V$, is proportional to number of particles $n$, we can see that the Krylov complexity is also proportional to volume. This matches to the complexity equals volume conjecture in the context of AdS/CFT. 

We can also easily obtain the expression for K-entropy $S_K$ defined in \cite{
Barbon:2019wsy} as
\begin{align}
\begin{split}
\label{eq:entanglementEntropy}
S_K &=  - \sum_{n = 0}^\infty  |\phi_n|^2 \text{ ln} |\phi_n|^2\\
    &= - \sum_{n = 0}^\infty \frac{\text{tanh}^{2n}r_k}{\text{cosh}^2r_k} \text{ln}\frac{\text{tanh}^{2n}r_k}{\text{cosh}^2r_k} \\
    &= -\sum_{n = 0}^\infty \frac{\text{tanh}^{2n}r_k}{\text{cosh}^2r_k}  \big({\text{ln}(\text{tanh}^{2n}r_k) - \text{ln}(\text{cosh}^{2}r_k)  \big) }\\
    &= \text{ln(cosh}^2r_k) \text{cosh}^2r_k - \text{ln(sinh}^2r_k) \text{sinh}^2r_k
\end{split}
\end{align}

\section{\textcolor{Sepia}{\textbf{ \Large Application to de Sitter Cosmology}}}\label{sec:Desitter}
Now that we have studied the Krylov complexity and chaos for the general cosmolgical perturbations, we will apply it in the context of  exponentially expanding de Sitter background. There are several motivations in choosing the de Sitter background in particular. Usually obtaining analytical solutions for (\ref{eq:evolutions}) is a difficult task and have to rely on numerical techniques, however for de Sitter background, there exists an exact expression for squeezing parameters $r_k$ and $\phi_k$ which makes it easier to study chaos and complexity. While this is one reason for this choice, the more important reason is that original motivation to study complexity and chaos comes from various conjectures in AdS/CFT. Since the universe  we live in is de Sitter in naure rather than Anti- de Sitter, checking these conjectures for the de Sitter background is also an equally important task. This has motivated us to study complexity and chaos in this space. 

For an exponentially expanding de Sitter background, the scale factor $a(\tau)$ is given by
\begin{equation}
\label{eq:scalefactor}
    a(\tau) = \frac{-1}{H\tau}
\end{equation}
where $-\infty < \tau < 0$ so that $z'/z = -1/\tau$.
For effective sound speed $c_s=1$, we can obtain the exact solution to squeezing parameters to the differential equations (\ref{eq:evolutions}). However, for other effective sound speeds, we will rely on the numerical methods. 
 \subsection{\textcolor{Sepia}{\textbf{ \Large Effective sound speed: $c_s = 1$}}} 
For $c_s=1$, the exact solutions to squeezing parameters to the differential equations (\ref{eq:evolutions}) are 
 \begin{align}
    r_k(\tau) &= -\sinh^{-1}\left( \frac{1}{2k\tau} \right) \nonumber\\
    \phi_k(\tau) &= -\frac{\pi}{4} - \frac{1}{2}\tan^{-1}\left( \frac{1}{2k\tau} \right) 
 \end{align}
During early times $k |\tau| \gg 1 $, so the modes are inside the horizon and squeezing parameters $r_k$ is almost zero $r_k \approx  -1 / (2k\tau) \ll 1$. During this limit, squeezing angle is also constant at $\phi _k  \approx -\pi /4$. At late times $k |\tau| \ll 1 $, the modes are outside the horizon. During this limit, the system behaves like an inverted harmonic oscillator and the squeezing grows with time $r_k \approx |ln(-k \tau)| \gg 1$
 
 Given the expression for the squeezing parameters,  we can now obtain an exact expressions of Krylov complexity $K$, Lancoz coefficients $b_n$, and Lyapunov exponents for de Sitter space with $c_s =1$ as
\begin{align}
    K &= \sinh^2r_k = \frac{1}{4k^2\tau^2} \\
    b_n &=  \frac{-z'}{2z}n =  \frac{n}{2 \tau} \\
    \lambda &=  \frac{-z'}{z} =  \frac{1}{\tau}
\end{align}
During early times $k |\tau| \gg 1 $, Krylov complexity is almost zero, $K \approx 0$, as expected. Similarly, Lyapunov exponent also has a very low value $\lambda \ll 1$ during this limit. For late times, $k |\tau| \ll 1 $ Krylov complexity grows exponentially $ K \gg 1$ and Lyapunov exponent is also much larger $\lambda \gg 1$. This is a strong feature of a chaotic system. It is interesting to see that Krylov complexity doesn't saturate in time, and rather keeps on increasing. This has to do with the fact that the expression for scale factor (\ref{eq:scalefactor}) is time dependent. So, the increase in chaotic nature of the de-Sitter background is primarily due to the time dependence of scale factor. 

\begin{figure*}[htb!]
	\centering
	\includegraphics[width=15cm,height=10cm]{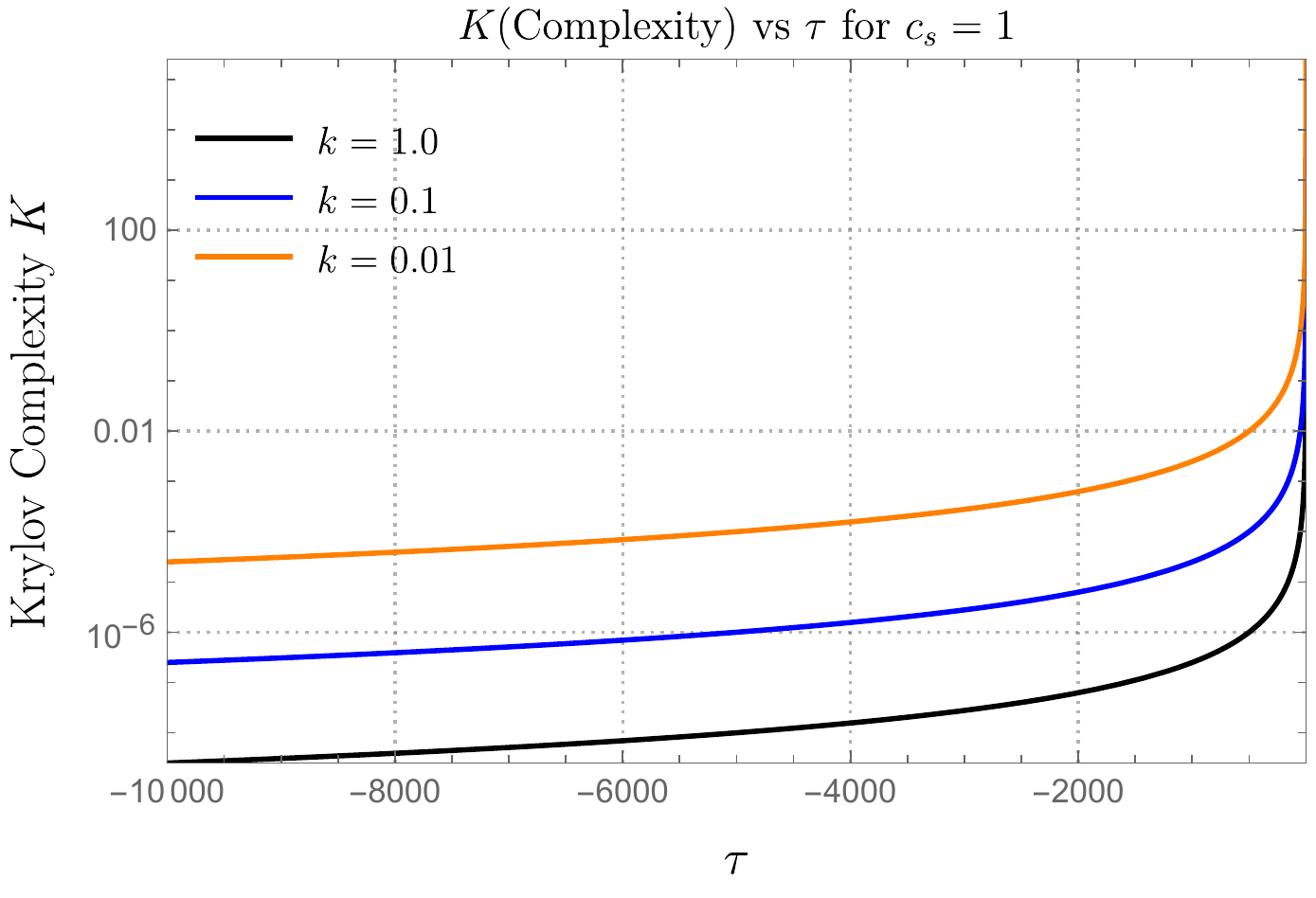}
	\caption{Krylov complexity as a function of conformal time $\tau$ for exponentially expanding de Sitter universe with different wave numbers $k$. Krylov complexity grows exponentially with $\tau$ which is a sign of chaotic system }
	\label{fig:krylov}
\end{figure*}

\begin{figure*}[htb!]
	\centering
	\includegraphics[width=15cm,height=10cm]{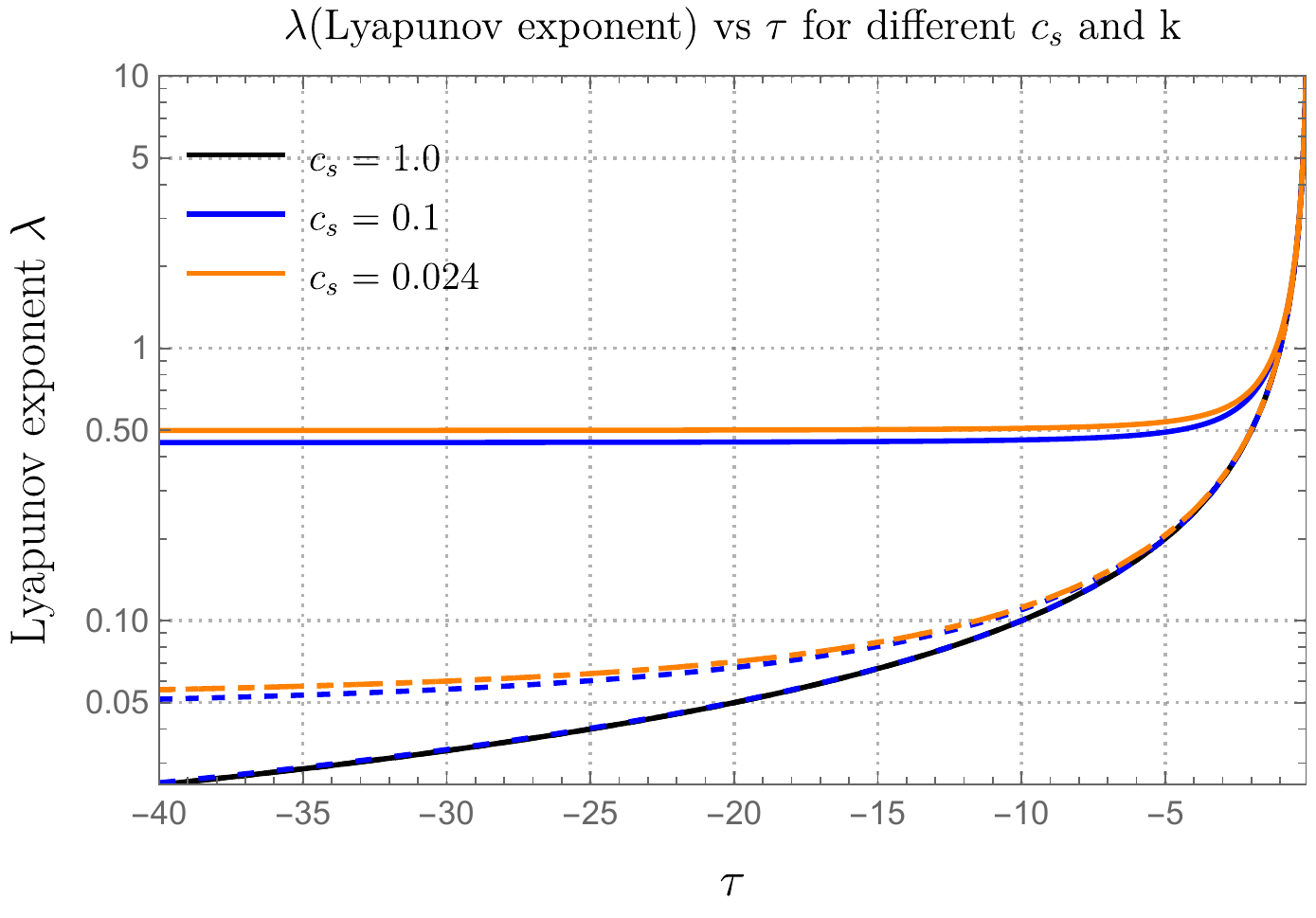}
	\caption{Lyapunov exponent as function of conformal time $\tau$ for different values of $c_s$ and k. For each color, solid line belongs to $k = 1$, dashed line to $k = 0.1$ and dashed medium to $k = 0.01$.}
	\label{fig:lyapunov}
\end{figure*}

\begin{figure*}[htb!]
	\centering
	\includegraphics[width=15cm,height=10cm]{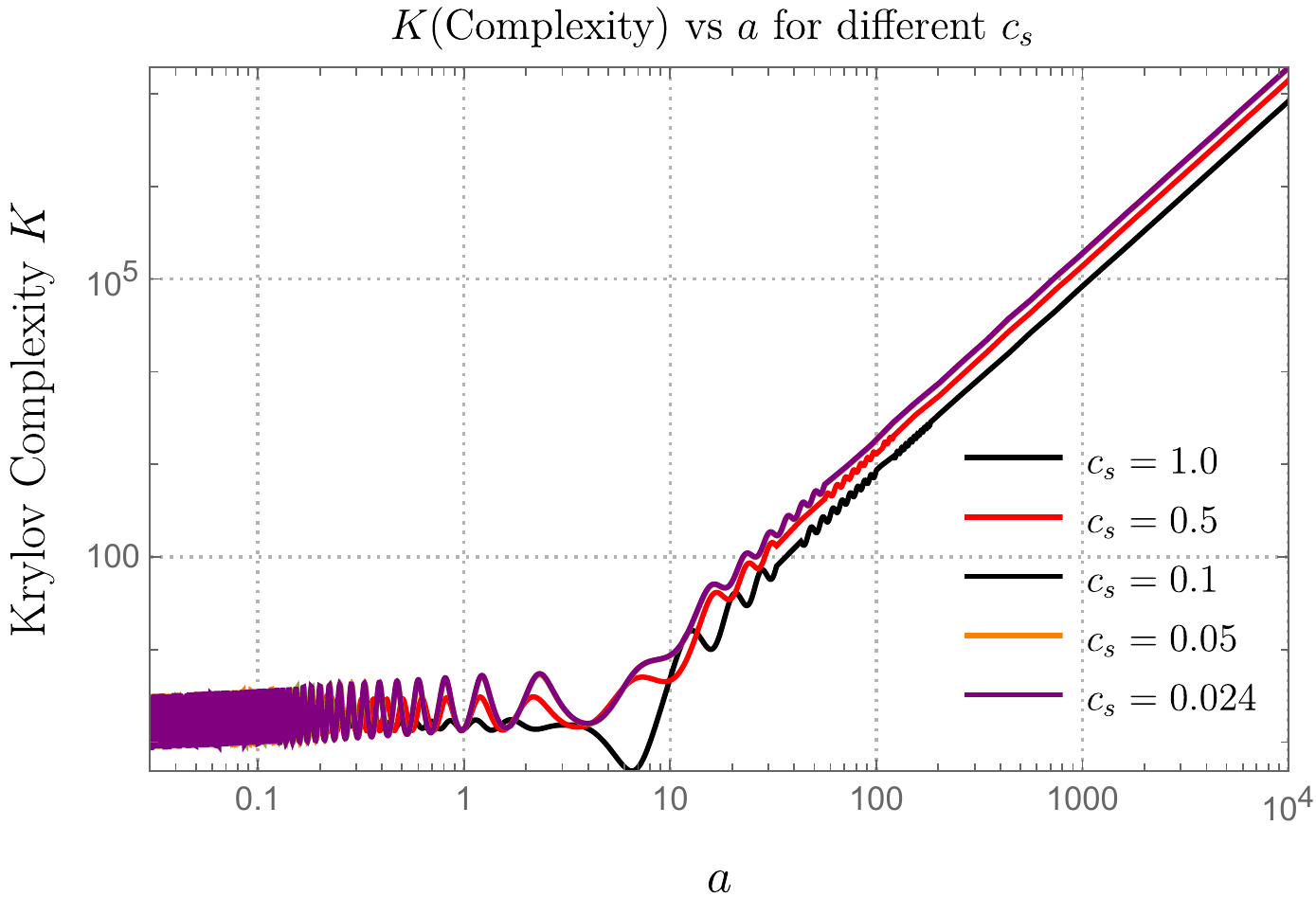}
	\caption{Krylov complexity as function of scale factor $a$ for different values of $c_s$}
	\label{fig:krylovgeneral}
\end{figure*}

In figure \ref{fig:krylov}, we have plotted the Krylov complexity for different wave numbers as a function of conformal time $\tau$ for $c_s = 1$. Krylov complexity grows exponentially with $\tau$ showing that the system is chaotic in nature. During early times, complexity is inversely proportional to the wave number while for late times the difference due to wave number is less. This exponential growth in complexity signifies chaos which is captured by lyapunov exponent in figure \ref{fig:lyapunov}.

We can also make an comparision of the Krylov complexity with Nielsen's geometric complexity. In \cite{Bhattacharyya:2020rpy}, the geometric complexity for cosmological perturbations was obtained to be
\begin{equation}
\begin{aligned}
 C &= \left|\text{ln}\left| \frac{1+\text{exp}(-2i\phi_k(\tau))\text{tanh}r_k(\tau)}{1-\text{exp}(-2i\phi_k(\tau))\text{tanh}r_k(\tau)}\right|\right| \\
 & ~+
 \left| \text{tanh}^{-1}(\text{sin}(2\phi_k(\tau))\text{sinh}(2r_k(\tau))) \right| 
\end{aligned}
\end{equation}
From hindsight, it looks like this measure of complexity captures the physics of the system in more detail than Krylov's complexity. In particular, geometric complexity is dependent on the squeezing angle too while Krylov's complexity is not. But one has to understand that Nielsen's geometric complexity has lots of ambiguities such as arbitrary choices of gates, reference and target states. This makes it very difficult to study in the context of cosmological evolution. Nielsen's measure is a good approach while constructing optimal quantum circuit in lab, but a difficult choice for cosmology and holography. 

In \cite{Maldacena:2015waa}, quantum chaos was conjectured to be bounded from above by the temperature of the system. We can also relate the lyapunov exponent we have obtained with the temperatue. In particular, temperature of the expanding universe $T$, is related to Hubble constant
by 
\begin{equation}
  T \approx H/2\pi.  
\end{equation}
 For de-Sitter background at time $\tau_0$
\begin{equation}
    |\tau_0 |= 1/H_{dS}
\end{equation}
Then, we obtain the Lyapunov exponent at $\tau_0 $ as
\begin{equation}
    \lambda_{\tau_0} = 2 \pi T
\end{equation}
Therefore, chaos in de-Sitter space saturates the bound conjectured in \cite{Maldacena:2015waa}. Interestingly this bound on cosmological complexity also matches with previous results obtained via other complexity measures \cite{Bhattacharyya:2020kgu,Bhattacharyya:2020rpy}.

Finally, we can also give an expression for K-entropy (\ref{eq:entanglementEntropy})
\begin{equation}
    S_K = \ln{\left(1+\frac{1}{4k^2\tau^2}\right)}\left(1+\frac{1}{4k^2\tau^2}\right) - \ln{\left(\frac{1}{4k^2\tau^2}\right)}\frac{1}{4k^2\tau^2}
\end{equation}
During early times, the difference between complexity and $K-$ entropy is less but for late times, the difference is huge. This shows that, complexity can grow even after system has achieved saturation. 

 \subsection{\textcolor{Sepia}{\textbf{ \Large Effective sound speed:  $0.024 \leq c_s \leq 1$}}}
 For other effective sound speed, we can give an exact expression Lancoz coefficients $b_n$, and Lyapunov exponents:
 \begin{align}
    b_n &= \frac{n}{2} \left( \sqrt{\left(\frac{k}{2}(1-c_s^2) \right)^2+ \left(\frac{1}{\tau} \right)^2} \right)\\
    \lambda &=  \sqrt{\left(\frac{k}{2}(1-c_s^2) \right)^2+ \left(\frac{1}{\tau} \right)^2}
\end{align}
Interestingly unlike the case for $c_s =1$, the lyapunov exponent is also dependent on the mode vectors $k$. Since Lancoz coefficients grows linearly with $n$, the system is chaotic too. In \ref{fig:lyapunov}, we have plotted Lyapunov exponent for values of $k = 1, 0.1, 0.01$ and $c_s = 1.0, 0.1, 0.024$. The lyapunov exponent is bounded from below by $c_s = 1$. For other $c_s$, we can see that lyapunov exponent is strongly dependent on $k$. For example, lyapunov exponent for $c_s = 0.1$ with $k = 1$ is significantly higher than for $c_s = 0.024$ with $k = 0.01$.

For computing Krylov complexity, we will rely on the numerical tools to obtain the solutions to squeezing parameters for differential equations (\ref{eq:evolutions}). 
 In order to make the numerical solution easier, instead of using the conformal time $\tau$ as the dynamical variable as in (\ref{eq:evolutions}), we will perform the change in variable from $\tau$ to $a(\tau)$
\begin{equation}
    \tau\longrightarrow a(\tau):~~~~\frac{d}{d\tau}=\frac{d}{da(\tau)}\frac{da(\tau)}{d\tau}=a'(\tau)\frac{d}{da(\tau)}
\end{equation} 
Consequently,  differential equations (\ref{eq:evolutions}) can be recast in terms of  $a(\tau)$ as:
\begin{align} \label{eq:diffeqnswa1}
&\frac{dr_k(a)}{da} = -\frac{\beta_k(a)}{a'} \cos(2(\tilde{\phi_k(a)}-\phi_k(a))),\\
\label{eq:diffeqnswa2}
&\frac{d\phi_k(a)}{da} = \frac{\Omega_k}{a'} -\frac{\beta_k(a)}{a'} \coth2 r_k(a)\sin(2(\tilde{\phi_k(a)}-\phi_k(a)))
\end{align} 

For numerical solutions, we will also fix the boundary conditions at late time scale $\tau = \tau_0$ where $a(\tau_0)= 1$, and the squeezing parameters are fixed to be $r_k(a(\tau_0)) = \phi_k(a(\tau_0)) = 1$. In Fig. \ref{fig:krylovgeneral}, we have plotted Krylov complexity as a function of scale factor for different values of effective sound speed with the solutions of squeezing parameters obtained numerically. The mode vector $k$ is fixed to be $1$ for all sound speeds. Like Lyapunov exponent, we can see that complexity is bounded from below by $c_s = 1.0$.

\section{\textcolor{Sepia}{\textbf{ \Large Conclusion}}}\label{sec:Conclusion}
In our work, we studied Krylov complexity and chaos for cosmological perturbations using squeezed states formalism and applied it to the de Sitter background. The main conclusions are as follows:
\begin{itemize}
    \item We have obtained an explicit relation for Krylov complexity and chaos for cosmological perturbations
    \begin{equation}
        K = \sinh^2{r_k}, \lambda = \sqrt{\left(\frac{k}{2}(1-c_s^2) \right)^2+ \left(\frac{1}{\tau} \right)^2} \nonumber
    \end{equation}
    Interesting, Krylov complexity is equal to averal particle number in each mode. Since volume is propotional to number of particles, the Krylov complexity is also propotional to volume. 
    \item For de Sitter background with $c_s = 1$, the expressions for complexity and chaos are
\begin{align}
    K &= \sinh^2r_k = \frac{1}{4k^2\tau^2} \nonumber \\
    b_n &=  \frac{-z'}{2z}n =  \frac{n}{2 \tau} \nonumber \\
    \lambda &=  \frac{-z'}{z} =  \frac{1}{\tau} \nonumber
\end{align}
This lyapunov exponent can also be written as $\lambda = 2\pi T$ which saturates the bound conjectured in \cite{Maldacena:2015waa}. For other sound speed, we rely on numerical tools and found that both Krylov complexity and lyapunov exponent are bounded from below by values for $c_s = 1$. 
    
\end{itemize}
In this work, we mainly focused on de Sitter background because of it's simplicity as well as it's wide applications. There are several other interesting and realistic cosmological backgrounds such as inflation, radiation dominated models, cosmological islands where these concepts can be further explored. Our work can also be seen as studying complexity and chaos on quantum fields in curved background. These concepts can be further explored in quantum fields and in holography. We saw that for squeezed states formalism, Krylov complexity is proportional to volume of the system. It would be interesting to see if it has any relevance to "Complexity = Volume" conjectures in hologrphy.  

\textbf{Acknowledgement:}
~~~The Visiting Post Doctoral research fellowship of SC is supported by the J. C. Bose National Fellowship of Director, Professor Rajesh Gopakumar, ICTS, TIFR, Bengaluru. SC also would line to thank ICTS, TIFR, Bengaluru for providing the work friendly environment. SC also thank all the members of our newly formed virtual international non-profit consortium Quantum Structures of the Space-Time \& Matter (QASTM) for for elaborative discussions. KA would like to thank the members of the QASTM Forum for useful discussions. Last but not least, we would like to acknowledge our debt to the people belonging to the various part of the
world for their generous and steady support for research
in natural sciences.

\bibliography{thebibliography}
\bibliographystyle{utphys}

\end{document}